\documentclass[final,3p,times]{elsarticle}
\usepackage{amsfonts}
\usepackage{amssymb}
\usepackage{amsmath}
\usepackage{graphics}
\usepackage{epsfig}
\usepackage{epstopdf}
\usepackage{amsthm}
\usepackage{textcomp}
\usepackage{color}
\usepackage{float}
\usepackage{multirow}
\usepackage{caption}
\usepackage{mathrsfs}
\usepackage{float}
\biboptions{sort&compress}

\journal{Security and Communication Networks}

\begin{document}

\begin{frontmatter}

\title{Defending against the advanced persistent threat: An optimal control approach}

\cortext[cor1]{Corresponding author}
%%\cortext[cor2]{Principal corresponding author}

\author[label1,label2]{Pengdeng Li}
\ead{1414797521@qq.com}

\author[label1,label2]{Xiaofan Yang\corref{cor1}}
\ead{xfyang1964@gmail.com}

\author[label1,label2]{Qingyu Xiong}
\ead{Xiong03@equ.edu.cn}

\author[label1,label2]{Junhao Wen}
\ead{jhwen@cqu.edu.cn}

\author[label3]{Yuan Yan Tang}
\ead{yytang@umac.mo}

\address[label1]{Key Laboratory of Dependable Service Computing in Cyber Physical Society, Ministry of Education, Chongqing University, Chongqing, 400044, China}

\address[label2]{School of Software Engineering, Chongqing University, Chongqing, 400044, China}

\address[label3]{Department of Computer and Information Science, The University of Macau, Macau}

\begin{abstract}
%% Text of abstract

The new cyber attack pattern of advanced persistent threat (APT) has posed a serious threat to modern society. This paper addresses the APT defense problem, i.e., the problem of how to effectively defend against an APT campaign. Based on a novel APT attack-defense model, the effectiveness of an APT defense strategy is quantified. Thereby, the APT defense problem is modeled as an optimal control problem, in which an optimal control stands for a most effective APT defense strategy. The existence of an optimal control is proved, and an optimality system is derived. Consequently, an optimal control can be figured out by solving the optimality system. Some examples of the optimal control are given. Finally, the influence of some factors on the effectiveness of an optimal control is examined through computer experiments. These findings help organizations to work out policies of defending against APTs.

\end{abstract}


\end{frontmatter}

%%
%% Start line numbering here if you want
%%
% \linenumbers

%% main text

\section{Introduction}

Nowadays, the daily operation of most organizations, ranging from large enterprises and financial institutions to government sectors and military branches, depends largely on computers and networks. However, this dependency renders the organizations vulnerable to a wide range of cyber attacks. Traditional cyber attacks include computer viruses, worms and spyware. Conventional cyber defense measures including firewall and intrusion detection turn out to be effective in withstanding these cyber attacks \cite{Kostopoulos2012, Singer2014}.    

The cyber security landscape has changed drastically over the past few years. A new type of cyber attack --- advanced persistent threat (APT) --- has posed an unprecedentedly serious threat to modern society. According to report, many high-profile organizations have experienced APTs \cite{Virvilis2013}, and the number of APTs has been increasing rapidly \cite{Rass2017}. Compared with traditional cyber attacks, APTs exhibit two distinctive characteristics: (a) The attacker of an APT is a well-resourced and well-organized group, with the goal of stealing as many sensitive data as possible from a specific organization. (b) Based on meticulous reconnaissance, the attacker is going to launch a preliminary advanced social engineering attack on a few target users to gain footholds in the organization and then to gain access to critical information stealthily and slowly \cite{Tankard2011, Cole2013, Wrightson2015}. Due to these characteristics, APTs can evade traditional detection, causing tremendous damage to organizations. To date, the detection of APTs is far from mature \cite{Friedberg2015, Marchetti2016}. Consequently, the APT defense problem, i.e., the problem of how to effectively defend against APTs, has become a major concern in the field of cybersecurity. 

As a branch of applied mathematics, optimal control theory aims to solve a class of optimization problems in which, subject to a set of dynamic constraints, we seek to find a function (control) so that an objective functional is optimized \cite{Kirk2004, Liberzon2012}. In real world applications, the set of dynamic constraints represents a dynamic environment, a control represents a time-varying strategy, and the objective functional represents an index to be maximized or minimized. Optimal control theory has been successfully applied to some aspects of cybersecurity \cite{Khouzani2010, Khouzani2011, RenJG2013, ChenLJ2015, YangLX2016, Nowzari2016, ZhangTR2017, LiuWP2017}. To our knowledge, the APT defense problem has yet to be addressed in the framework of optimal control theory. To model the problem as an optimal control problem, we have to formulate an APT defense strategy  as a control, characterize the state evolution of an organization as a set of dynamic constraints, and quantify the effectiveness of an APT defense strategy as an objective functional. The key to the modeling process is to accurately characterize the state evolution of an organization by employing the epidemic modeling technique \cite{Britton2003}. 

Individual-level epidemic models refer to epidemic models in which the state evolution of each individual in a population is characterized by one or a few separate differential equations. As compared with the coarse-fined state-level epidemic models \cite{Piqueira2009, FengLP2012, Mishra2013, YaoY2013, FengLP2015, RenJG2017} and the intermediate degree-level epidemic models \cite{Satorras2001, Satorras2002, Castellano2010, YangLX2013, RenJG2016, LiuWP2016, YangLX2017f}, the fine-coarsed individual-level epidemic models can characterize spreading processes more accurately, because they can perfectly accommodate the network topology. The individual-level epidemic modeling technique has been successfully applied to areas such as the epidemic spreading \cite{Mieghem2009, Mieghem2011, Sahneh2012, Sahneh2013}, the malware spreading \cite{XuSH2012a, XuSH2012b, XuSH2014a, YangLX2017a, YangLX2017b, YangLX2017c}, and the rumor spreading \cite{YangLX2017d}. In particular, a number of APT attack-defense models have recently been proposed by employing this technique \cite{XuSH2015, ZhengR2015, YangLX2017e, ZhengR2016}. 

This paper focuses on the APT defense problem. Based on a novel individual-level APT attack-defense model, the effectiveness of an APT defense strategy is quantified. On this basis, the APT defense problem is modeled as an optimal control problem, in which an optimal control represents a most effective APT defense strategy. The existence of an optimal control to the optimal control problem is proved, and an optimality system for the optimal control problem is derived. Therefore, an optimal control can be figured out by solving the optimality system. Some examples of the optimal control are presented. Finally, the influence of some factors on the effectiveness of an optimal control is examined through computer simulations. To our knowledge, this is the first time the APT defense problem is dealt with in this way. These findings help organizations to work out policies of defending against APTs.

The remaining materials are organized in this fashion. Section 2 models the APT defense problem as an optimal control problem. Section 3 studies the optimal control problem. Some most effective APT defense strategies are given in Section 4. Section 5 discusses the influence of different factors on the optimal effectiveness. This work is closed by Section 6.

\newtheorem{rk}{Remark}
\newproof{pf}{Proof}
\newtheorem{thm}{Theorem}
\newtheorem{lm}{Lemma}
\newtheorem{exm}{Example}
\newtheorem{cor}{Corollary}
\newtheorem{de}{Definition}
\newtheorem{cl}{Claim}
\newtheorem{pro}{Proposition}
\newtheorem{con}{Conjecture}

\newproof{pfcl1}{Proof of Claim 1}
\newproof{pfcl2}{Proof of Claim 2}

\section{The modeling of the APT defense problem}

The goal of this paper is to solve the following problem:

\emph{The APT defense problem:} Defend an organization against APTs in an effective way.

To achieve the goal, we have to model the problem. The modeling process consists of the following four steps:

\emph{Step 1:} Introduce preliminary terminologies and notations.

\emph{Step 2:} Establish an APT attack-defense model.

\emph{Step 3:} Quantify the effectiveness of an APT defense strategy.

\emph{Step 4:} Model the APT defense problem as an optimal control problem.
 
Now, let us proceed by following this four-step procedure.

\subsection{Preliminary terminologies and notations} 

Consider an organization with a set of $N$ computer systems labeled $1, 2, \cdots, N$. Let $G = (V, E)$ denote the access network of the organization, where (a) each node stands for a system, i.e., $V = \{1, 2, \cdots, N\}$, and (b) $(i, j) \in E$ if and only if system $i$ has access to system $j$. Let $\mathbf{A} = \left[a_{ij}\right]_{N \times N}$ denote the adjacency matrix for the network, i.e., $a_{ij} = 1$ or 0 according as $(i, j) \in E$ or not. 

Suppose an APT campaign to the organization starts at time $t = 0$ and terminates at time $t = T$. Suppose at any time $t \in [0, T]$ every node in the organization is either \emph{secure}, i.e. under the defender's control, or \emph{compromised}, i.e., under the attacker's control. Let $X_i(t) = 0$ and 1 denote the event that node $i$ is secure and compromised at time $t$, respectively.
The vector
\begin{equation}
  \mathbf{X}(t) = \left[X_1(t), X_2(t), \cdots, X_N(t) \right]
\end{equation}
stands for the \emph{state} of the organization at time $t$. Let $S_i(t)$ and $C_i(t)$ denote the probability of the event that node $i$ is secure and compromised at time $t$, respectively. That is,
\begin{equation}
  S_i(t) = \Pr\left\{X_i(t) = 0\right\}, \quad C_i(t) = \Pr\left\{X_i(t) = 1\right\}.
\end{equation}
As $S_i(t)+C_i(t)\equiv 1$, the vector
\begin{equation}
  \mathbf{C}(t) = \left[C_1(t), \cdots, C_N(t)\right]^T
\end{equation}
stands for the \emph{expected state} of the organization at time $t$.

From the attacker's perspective, each secure node in the organization is subject to the external attack. Let $a_i$ denote the cost per unit time for attacking a secure node $i$. The vector
\begin{equation}
  \mathbf{a} = \left[a_1, \cdots, a_N\right]
\end{equation}
stands for an \emph{attack strategy}. Additionally, each secure node is vulnerable to all the neighboring compromised nodes. 

From the defender's perspective, each secure node in the organization is protected from being compromised. Let $x_i(t)$ denote the cost per unit time for protecting the secure node $i$ at time $t$. The vector-valued function
\begin{equation}
  \mathbf{x}(t) = \left[x_1(t), \cdots, x_N(t)\right], \quad 0 \leq t \leq T
\end{equation}
stands for a \emph{prevention strategy}. Additionally, each compromised node in the organization is revovered. Let $y_i(t)$ denote the cost per unit time for recovering the compromised node $i$ at time $t$. The vector-valued function
\begin{equation}
\mathbf{y}(t) = \left[y_1(t), \cdots, y_N(t)\right], \quad 0 \leq t \leq T
\end{equation}
stands for a \emph{recovery strategy}. We refer to the vector-valued function 
\begin{equation}
  \mathbf{u}(t) = \left[\mathbf{x}(t), \mathbf{y}(t)\right] = \left[x_1(t), \cdots, x_N(t), y_1(t), \cdots, y_N(t)\right], \quad 0 \leq t \leq T
\end{equation}
as an \emph{APT defense strategy}.

\subsection{An APT attack-defense model}

For fundamental knowledge on differential dynamical systems, see Ref. \cite{Robinson2004}. For our purpose, let us impose a set of hypotheses as follows.

\begin{enumerate}
	
	\item [(H$_1$)] Due to the external attack and prevention, a secure node $i$ gets compromised at time $t$ at the average rate $\frac{a_i}{x_i(t)}$. The rationality of this hypothesis lies in that the average rate is proportional to the attack cost per unit time and is inversely proportional to the prevention cost per unit time. 
	
	\item [(H$_2$)] Due to the internal infection and prevention, a secure node $i$ gets compromised at time $t$ at the average rate $\frac{\beta\sum_{j=1}^{N}a_{ji}C_j(t)}{x_i(t)}$, where $\beta > 0$ is a constant, which we refer to as the \emph{infection force}. The rationality of this assumption lies in that the average rate is proportional to the probability of each neighboring node being compromised and is inversely proportional to the prevention cost per unit time.
	
	\item [(H$_3$)] Due to the recovery, a compromised node $i$ becomes secure at time $t$ at the average rate $y_i(t)$. The rationality of this assumption lies in that the average rate is proportional to the recovery cost per unit time.
	
\end{enumerate}

According to these hypotheses, the state transitions of a node are shown in Fig. 1. Hence, the time evolution of the expected state of the organization obeys the following dynamical system:
\begin{equation}
	\frac{dC_i(t)}{dt}= \frac{1}{x_i(t)}\left[a_i + \beta\sum_{j=1}^{N}a_{ji}C_j(t)\right]\left[1 - C_i(t)\right] - y_i(t)C_i(t), \quad 0 \leq t \leq T, i = 1, \cdots, N.
\end{equation}

\noindent We refer to the model as the \emph{APT attack-defense model}. 

\begin{figure}[H]
	\setlength{\abovecaptionskip}{0.cm}
	\setlength{\belowcaptionskip}{-0.cm}
	\centering
	\includegraphics[width=0.6\textwidth]{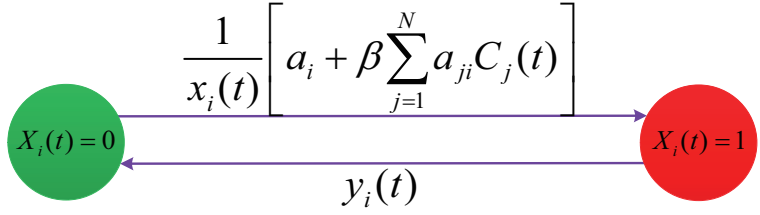}
	\caption{The diagram of state transitions of a node under the hypotheses (H$_1$)-(H$_3$).}
\end{figure}

The APT attack-defense model can be written in matrix-vector notation as
\begin{equation}
  \frac{d\mathbf{C}(t)}{dt} = \mathbf{F}(\mathbf{C}(t),\mathbf{u}(t)), \quad 0 \leq t \leq T.
\end{equation}

\subsection{The effectiveness of an APT defense strategy}

The defender's goal is to find the most effective APT defense strategy. To achieve the goal, we have to quantify the effectiveness of an APT defense strategy. For this purpose, let us introduce an additional set of hypotheses as follows.

\begin{enumerate}
	
\item[(H$_4$)] The prevention cost per unit time is bounded from above by $\overline{x}$ and from below by $\underline{x} > 0$, and the recovery cost per unit time is bounded from above by $\overline{y}$ and from below by $\underline{y} > 0$. That is, the admissible set of APT defense strategies is given by
\begin{equation}
  \mathscr{U}=\left\{\mathbf{u}\in \left(L^2[0,T]\right)^{2N} \mid  0 < \underline{x} \leq x_i(t) \leq \overline{x}, 0 < \underline{y} \leq y_i(t) \leq \overline{y}, 0 \leq t \leq T, 1 \leq i \leq N \right\},
\end{equation}
where $L^2[0,T]$ denote the set of all the Lebesgue square integrable functions defined on the interval $[0,T]$ \cite{Stein2005}.

\item[(H$_5$)] The amount of losses caused by a compromised node $i$ in the infinitesimal time interval $[t, t + dt]$ is $w_idt$, where $w_i = \sum_{j=1}^Na_{ij}$ stands for the out-degree of node $i$ in the network. The rationality of this hypothesis lies in that the more nodes a node has access to, the more serious the consequence when it is compromised \cite{Kempe2005, ChenW2009}.
	
\end{enumerate}

According to the hypotheses, the expected loss of the organization in the time horizon $[0, T]$ when implementing an APT defense strategy $\mathbf{u} = \left[\mathbf{x}, \mathbf{y}\right]$ is 
\begin{equation}
  Loss(\mathbf{u}) = \sum_{i=1}^N \int_0^T w_iC_i(t)dt, 
\end{equation}
and the overall cost for implementing the APT defense strategy is 
\begin{equation}
  Cost(\mathbf{u}) = \sum_{i=1}^N \int_0^T \left[x_i(t)+y_i(t)\right]dt. 
\end{equation}
Hence, the effectiveness of the APT defense strategy $\mathbf{u}$ can be 
measured by the quantity
\begin{equation}
  J(\mathbf{u}) = Loss(\mathbf{u}) + Cost(\mathbf{u}) = \sum_{i=1}^N \int_0^T w_iC_i(t)dt + \sum_{i=1}^N \int_0^T \left[x_i(t)+y_i(t)\right]dt.
\end{equation}
Obviously, the smaller this quantity, the more effective the APT defense strategy. Let 
\begin{equation}
L(\mathbf{C}(t),\mathbf{u}(t)) = \sum_{i=1}^N [w_i C_i(t) + x_i(t) + y_i(t)].
\end{equation}
Then
\begin{equation}
  J(\mathbf{u}) = \int_0^T L(\mathbf{C}(t),\mathbf{u}(t))dt.
\end{equation}

\subsection{The modeling of the APT defense strategy}

Based on the previous discussions, the APT defense problem boils down to the following optimal control problem:
\begin{equation}
\begin{split}
  &\min_{\mathbf{u}\in \mathscr{U}} J(\mathbf{u}) = \int_0^T L(\mathbf{C}(t),\mathbf{u}(t))dt, \\
  &\text{ subject to } \frac{d\mathbf{C}(t)}{dt} = \mathbf{F}(\mathbf{C}(t),\mathbf{u}(t)), \quad 0 \leq t \leq T, \\
  &\text{ } \quad\quad\quad\quad \mathbf{C}(0) = \mathbf{C}_0.
\end{split}
\end{equation}
Here, each control stands for an APT defense strategy, the objective functional stands for the effectiveness of an APT defense strategy, the set of constraints stands for the time evolution of the expected state of the organization, an optimal control stands for a most effective APT defense strategy, and the optimal value stands for the effectiveness of a most effective APT defense strategy.

\section{A theoretical analysis of the optimal control problem}

For fundamental knowledge on optimal control theory, see Refs. \cite{Kirk2004, Liberzon2012}. This section is devoted to studying the optimal control problem (16).  

\subsection{The existence of an optimal control} 

As an optimal control to the problem (16) represents a most effective APT defense strategy, it is critical to show that the problem does have an optimal control. For this purpose, we need the following lemma \cite{Liberzon2012}.

\begin{lm}
Problem (16) has an optimal control if the following five conditions hold simultaneously.
\begin{enumerate}
  \item [(C$_1$)] $\mathscr{U}$ is closed and covex.
  \item [(C$_2$)] There is $\mathbf{u} \in \mathscr{U}$ such that the adjunctive dynamical system is solvable.	
  \item [(C$_3$)] $\mathbf{F}(\mathbf{C},\mathbf{u})$ is bounded by a linear function in $\mathbf{C}$.
  \item [(C$_4$)] $L(\mathbf{C},\mathbf{u})$ is convex on $\mathscr{U}$.
  \item [(C$_5$)] $L(\mathbf{C},\mathbf{u})$ $\geq c_1||\mathbf{u}||^\rho+c_2$ for some vector norm $||\cdot||$, $\rho >1,c_1>0$ and $c_2$.
\end{enumerate}
\end{lm}

Next, let us show that the five conditions in Lemma 1 indeed hold.

\begin{lm}
	The admissible set $\mathscr{U}$ is closed.
\end{lm}

\noindent \emph{Proof:} Let $\mathbf{u} = (x_1,\cdots,x_N,y_1, \cdots, y_N)$ be a limit point of $\mathscr{U}$,
$\mathbf{u}^{(n)} = \left(x_1^{(n)},\cdots,x_N^{(n)}, y_1^{(n)},\cdots,y_N^{(n)}\right), n=1, 2,\cdots,
$ be a sequence of points in $\mathscr{U}$ that approaches $\mathbf{u}$. As $\left(L^2[0,T]\right)^{2N}$ is complete, we have $\mathbf{u} \in \left(L^2(0,T)\right)^{2N}$.
Hence, the claim follows from the observation that
\begin{equation}
	\underline{x} \leq x_i(t)=\lim_{n\to \infty}x_i^{(n)}(t) \leq \overline{x}, \quad
	\underline{y} \leq y_i(t)=\lim_{n\to \infty}y_i^{(n)}(t) \leq \overline{y}, \quad 0 \leq t \leq T, 1\leq i \leq N.
\end{equation}

\begin{lm}
	The admissible set $\mathscr{U}$ is convex.
\end{lm}

\noindent \emph{Proof:} Let
$
	\mathbf{u}^{(1)} = \left(x_1^{(1)},\cdots,x_N^{(1)},y_1^{(1)},\cdots,y_N^{(1)}\right), \mathbf{u}^{(2)} = \left(x_1^{(2)},\cdots,x_N^{(2)},y_1^{(2)},\cdots,y_N^{(2)}\right) \in \mathscr{U}
$, $0<\eta<1$. As $\left(L^2[0,T]\right)^{2N}$ is a real vector space, we get
$
	(1-\eta)\mathbf{u}^{(1)}(t)+\eta \mathbf{u}^{(2)}(t) \in \left(L^2[0,T]\right)^{2N}.
$
So, the claim follows from the observation that
\begin{equation}
	\underline{x} \leq (1-\eta)x_i^{(1)}(t)+\eta x_i^{(2)}(t) \leq \overline{x}, \quad
	\underline{y} \leq (1-\eta)y_i^{(1)}(t)+\eta y_i^{(2)}(t) \leq \overline{y}, \quad 0 \leq t \leq T, 1\leq i \leq N.
\end{equation}

\begin{lm}
	There is $\mathbf{u} \in \mathscr{U}$ such that the associated adjunctive dynamical system is solvable.
\end{lm}

\noindent \emph{Proof:} Consider the adjunctive dynamical system 
\begin{equation}
	\frac{d\mathbf{C}(t)}{dt} =
	\mathbf{F}(\mathbf{C}(t),\overline{\mathbf{u}}), \quad 0 \leq t \leq T,
\end{equation}
where $\mathbf{u}(t) \equiv \overline{\mathbf{u}}=(\overline{x},\cdots,\overline{x}, \overline{y},\cdots,\overline{y})$. As $\mathbf{F}(\mathbf{C},\overline{\mathbf{u}})$ is continuously differentiable, the claim follows from the Continuation Theorem for Differential Dynamical Systems \cite{Robinson2004}.

\begin{lm}
	$\mathbf{F}(\mathbf{C},\mathbf{u})$ is bounded by a linear function in $\mathbf{C}$.
\end{lm}

\noindent \emph{Proof:} The claim follows from the observation that for $0 \leq t \leq T$, $i=1,2,\cdots,N$,
\begin{equation}
\begin{split}
	\frac{1}{x_i(t)}\left[a_i + \beta \sum_{j=1}^{N}a_{ji}C_j(t) \right][1 - C_i(t)] - y_i(t)C_i(t)
	& \leq \frac{1}{\underline{x}}\left[a_i + \beta \sum_{j=1}^{N}a_{ji}C_j(t) \right] - \underline{y}C_i(t).
\end{split}
\end{equation}

\begin{lm}
	$L(\mathbf{C},\mathbf{u})$ is convex on $\mathscr{U}$.
\end{lm}

\noindent \emph{Proof:} Let $\mathbf{u}^{(1)}, \mathbf{u}^{(2)} \in \mathscr{U}$, $0<\eta<1$. Then
\begin{equation}
	L\left(\mathbf{C},(1-\eta)\mathbf{u}^{(1)}+\eta \mathbf{u}^{(2)} \right) = (1-\eta)L\left(\mathbf{C},\mathbf{u}^{(1)}\right)
	+ \eta L\left(\mathbf{C},\mathbf{u}^{(2)}\right).
\end{equation}

\begin{lm}
	$L(\mathbf{C},\mathbf{u}) \geq \frac{1}{\max\{\overline{x}, \overline{y}\}}||\mathbf{u}||_2^2$.
\end{lm}

\noindent \emph{Proof:} We have
\begin{equation}
	L(\mathbf{C},\mathbf{u})
	= \sum_{i=1}^N \left(w_i C_i+x_i+y_i\right)
	\geq  \sum_{i=1}^N \left(x_i+y_i \right) \geq \sum_{i=1}^N \left(\frac{x_i^2}{\overline{x}} + \frac{y_i^2}{\overline{y}} \right)  \geq \frac{1}{\max\{\overline{x}, \overline{y}\}}||\mathbf{u}||_2^2.
\end{equation}

Based on Lemmas 1-7, we get the following result.

\begin{thm}
	The problem (16) has an optimal control.
\end{thm}

This theorem guarantees that there is a most effective APT defense strategy.

\subsection{The optimality system}

It is known that the optimality system for an optimal control problem offers a method for numerically solving the problem. This subsection is intended to present the optimality system for the problem (16). For this purpose, consider the corresponding Hamiltonian
\[
	%\left\{
	\begin{aligned}
		H(\mathbf{C}(t),\mathbf{u}(t), \mathbf{\lambda}(t))
		&=\sum_{i=1}^N\left[w_i C_i(t) + x_i(t) + y_i(t)\right]\\
		& \quad +\sum_{i=1}^N\lambda_i(t)\left\{\frac{1}{x_i(t)}\left[a_i + \beta \sum_{j=1}^{N}a_{ji}C_j(t)\right][1 - C_i(t)] - y_i(t)C_i(t) \right\}.
	\end{aligned}
	%\right.
\]
where $\mathbf{\lambda} = (\lambda_1, \cdots, \lambda_N)^T$ is the adjoint.

\begin{thm}
Suppose $\mathbf{u}^*$ is an optimal control to the problem (16), $\mathbf{C}^*$ is the solution to the adjunctive dynamical system with $\mathbf{u}=\mathbf{u}^*$. Then, there exists $\mathbf{\lambda}^*$ with $\mathbf{\lambda}^*(T)=\mathbf{0}$ such that for $0 \leq t \leq T$, $1\leq i \leq N$,
\[
\left\{
\begin{aligned}
  &\frac{d\lambda_i^*(t)}{dt} = -w_i + y_i^*(t) \lambda_i^*(t) + \frac{\lambda_i^*(t)}{x_i^*(t)} \left[a_i + \beta \sum_{j=1}^{N} a_{ji} C_j^*(t)\right] - \beta \sum_{j=1}^{N} \frac{a_{ij} [1-C_j^*(t)] \lambda_j^*(t)}{x_j^*(t)},\\
  &x_i^*(t) = \max \left\{\min \left\{\left[\lambda_i^*(t) \left[a_i + \beta \sum_{j=1}^{N} a_{ji} C_j^*(t)\right] \left[1-C_i^*(t) \right] \right]^{\frac{1}{2}},\overline{x} \right\},\underline{x}\right\},\\
  &y_i^*(t) = \left\{
			           \begin{aligned}
			               \underline{y}, \quad \lambda_i^*(t) C_i^*(t) < 1, \\
			               \overline{y}, \quad \lambda_i^*(t) C_i^*(t) > 1, 
			           \end{aligned}
			           \right.
\end{aligned}
\right.
\]
\end{thm}

\noindent \emph{Proof:}
According to the Pontryagin Minimum Principle \cite{Kirk2004}, there exists $\mathbf{\lambda}^*$ such that
\[
	\frac{d\lambda_i^*(t)}{dt}=-\frac{\partial H(\mathbf{C}^*(t),\mathbf{u}^*(t),\mathbf{\lambda}^*(t))}{\partial C_i}, \quad 0\leq t \leq T, 1 \leq i \leq N
\]

\noindent Thus, the first $N$ equations in the claim follow by direct calculations. As the terminal cost is unspecified and the final state is free, the transversality condition $\mathbf{\lambda}^*(T)=\mathbf{0}$ holds. By using the optimality condition
\[
	\mathbf{u}^*(t)=\arg \min\limits_{\mathbf{u}\in \mathscr{U}} H(\mathbf{C}^*(t), \mathbf{u}(t), \mathbf{\lambda}^*(t)), \quad 0 \leq t \leq T.
\]
we get (a) either
\[
	\frac{\partial H(\mathbf{C}^*(t),\mathbf{u}^*(t),\mathbf{\lambda}^*(t))}{\partial x_i} = 1 - \frac{\lambda_i^*(t)}{\left( x_i^*(t) \right)^2} \left[a_i + \beta \sum_{i=1}^{N} a_{ji} C_j^*(t)\right] [1-C_i^*(t)] =0
\]
or $x_i^*(t)=\underline{x}$ or $x_i^*(t)=\overline{x}$, and (b)
\[
	\begin{aligned}
	y_i^*(t) &= \arg \min\limits_{\underline{y} \leq y_i(t) \leq \overline{y}} \left( 1 - \lambda_i^*(t) C_i^*(t) \right) y_i(t)
	= \left\{
	    \begin{aligned}
	        \underline{y}, \quad \lambda_i^*(t) C_i^*(t) < 1, \\
	        \overline{y}, \quad \lambda_i^*(t) C_i^*(t) > 1. 
	    \end{aligned}
	    \right.
	\end{aligned}
\]
Combining the above discussions, we get the optimality system for the problem (16) as follows.
\begin{equation}
	\left\{
	\begin{aligned}
		&\frac{dC_i(t)}{dt}= \frac{1}{x_i(t)}\left[a_i + \beta\sum_{j=1}^{N}a_{ji}C_j(t) \right][1 - C_i(t)] - y_i(t)C_i(t),\\
		&\frac{d\lambda_i(t)}{dt} = -w_i + y_i(t)\lambda_i(t) + \frac{\lambda_i(t)}{x_i(t)} \left[a_i + \beta \sum_{j=1}^{N} a_{ji} C_j(t) \right] - \beta \sum_{j=1}^{N} \frac{a_{ij} [1-C_j(t)] \lambda_j(t)}{x_j(t)},\\
		&x_i(t) = \max \left\{\min \left\{\left[\lambda_i(t) \left[a_i + \beta \sum_{j=1}^{N} a_{ji} C_j(t)\right] \left[1-C_i(t) \right] \right]^{\frac{1}{2}},\overline{x} \right\},\underline{x}\right\},\\
		&y_i(t) = \left\{
			        \begin{aligned}
			             \underline{y}, \quad \lambda_i(t) C_i(t) < 1, \\
			             \overline{y}, \quad \lambda_i(t) C_i(t) > 1, 
			        \end{aligned}
			        \right.
	\end{aligned}
	\right.
\end{equation}
where $\mathbf{C}(0) = \mathbf{C}_0$, $\lambda(T)=\mathbf{0}$, $0 \leq t \leq T, 1 \leq i \leq N$.

Applying the forward-backward Euler scheme to the optimality system, we can obtain an optimal control to the problem (16), i.e, a most effective APT defense strategy.

\section{Some most effective APT defense strategies}

In this section, we give some most effective APT defense strategies by solving the optimality system (23). For ease in observation, let us introduce two functions as follows. For an admissible control $\mathbf{u}$ to the problem (16), define the \emph{cumulative effectiveness (CE)} as
\begin{equation}
CE(t; \mathbf{u}) = \sum_{i=1}^N\int_0^t w_i C_i(s)ds + \sum_{i=1}^N\int_0^t \left[x_i(s) + y_i(s) \right]ds, \quad 0 \leq t \leq T,
\end{equation}
and define the \emph{superposed control (SC)} as
\begin{equation}
SC(t; \mathbf{u}) = \sum_{i=1}^N\left[x_i(t)+ y_i(t)\right], \quad 0 \leq t \leq T,
\end{equation}
Obviously, we have $CE(T; \mathbf{u}) = J(\mathbf{u})$.

For some optimal control problems, let us give the cumulative effectiveness and superposed control for an optimal control.  

\begin{exm}
Consider the problem (16) in which $G$ is a scale-free network with $N = 100$ nodes that is generated by executing the algorithm given in Ref. \cite{Barabasi1999}, $T=20$, $\beta = 0.001$, $\underline{x} = \underline{y} = 0.1$, $\overline{x} = \overline{y} = 0.7$, $a_i=0.1$, $0 \leq i \leq N$, and $C_i(0) = 0.1, 0 \leq i \leq N$. An optimal control to the problem is obtained by solving the optimality system (23). Fig. 2 plots the cumulative effectiveness and superposed control for the optimal control. For comparison purpose, the cumulative effectiveness and superposed control for three admissible static controls are also shown in Fig. 2.
\end{exm}

\begin{figure}[H]
	\setlength{\abovecaptionskip}{0.cm}
	\setlength{\belowcaptionskip}{-0.cm}
	\centering
	\includegraphics[width=0.8\textwidth]{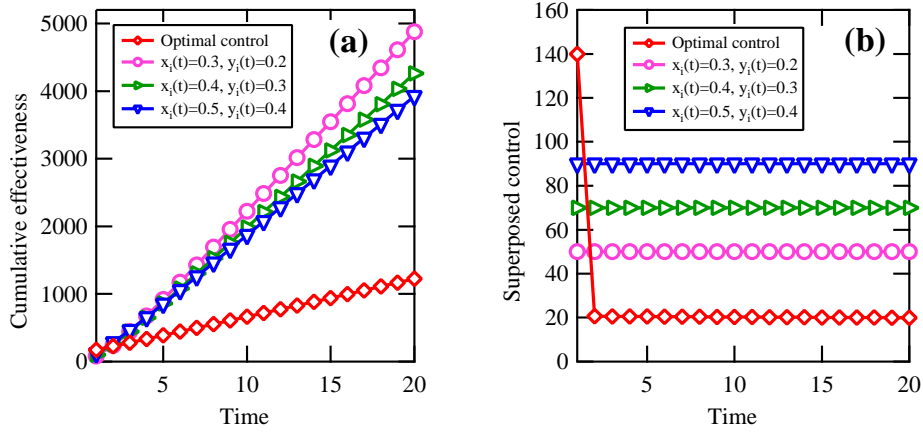}
	\caption{The cumulative effectiveness and superposed control for the optimal control and a few static controls in Example 1.}
\end{figure}

\begin{exm}
Consider the problem (16) in which $G$ is a small-world network with $N = 100$ nodes that is generated by executing the algorithm given in Ref. \cite{Watts1998}, $T=20$, $\beta = 0.001$, $\underline{x} = \underline{y} = 0.1$, $\overline{x} = \overline{y} = 0.7$, $a_i=0.1$, $0 \leq i \leq N$, and $C_i(0) = 0.1, 0 \leq i \leq N$. An optimal control to the problem is obtained by solving the optimality system (23). Fig. 3 depicts the cumulative effectiveness and superposed control for the optimal control. For comparison purpose, the cumulative effectiveness and superposed control for three admissible static controls are also shown in Fig. 3.
\end{exm}

\begin{figure}[H]
	\setlength{\abovecaptionskip}{0.cm}
	\setlength{\belowcaptionskip}{-0.cm}
	\centering
	\includegraphics[width=0.8\textwidth]{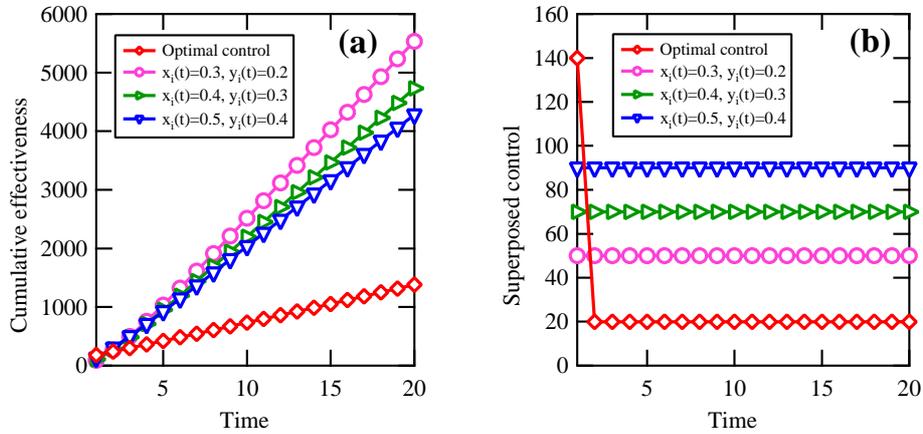}
	\caption{The cumulative effectiveness and superposed control for the optimal control and a few static controls in Example 2.}
\end{figure}

\begin{exm}
Consider the problem (16) in which $G$ is a realistic network given in Ref. \cite{KONECT}, $T=20$, $\beta = 0.001$, $\underline{x} = \underline{y} = 0.1$, $\overline{x} = \overline{y} = 0.7$, $a_i=0.1$, $0 \leq i \leq N$, and $C_i(0) = 0.1, 0 \leq i \leq N$. An optimal control to the problem is obtained by solving the optimality system (23). Fig. 4 exhibits the cumulative effectiveness and superposed control for the optimal control. For comparison purpose, the cumulative effectiveness and superposed control for three admissible static controls are also shown in Fig. 4.
\end{exm}

\begin{figure}[H]
	\setlength{\abovecaptionskip}{0.cm}
	\setlength{\belowcaptionskip}{-0.cm}
	\centering
	\includegraphics[width=0.8\textwidth]{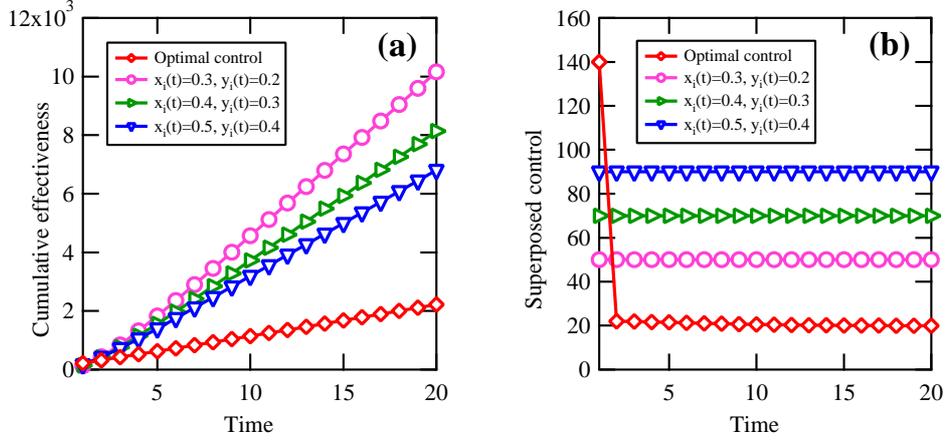}
	\caption{The cumulative effectiveness and superposed control for the optimal control and a few static controls in Example 3.}
\end{figure}

It is seen from the above three examples that a most effective APT defense strategy is significantly superior to any static APT defense strategy in terms of the effectiveness. This observation justifies our method. Additionally, the superposed control drops rapidly to a lower value.

\section{Further discussions}

This section is devoted to examining the influence of different factors on the optimal effectiveness of an admissible APT defense strategy. For ease in understanding these influences, let us introduce three quantities as follows. For an optimal control $\mathbf{u}^*$ to the problem (16), let $OL^*$, $OC^*$, and $OJ^*$ denote the corresponding expected loss, overall cost, and effectiveness, respectively. That is,
\begin{equation}
OL^* = Loss(\mathbf{u}^*), \quad OC^* = Cost(\mathbf{u}^*), \quad OJ^* = OL ^* + OC^* = J(\mathbf{u}^*).
\end{equation}

\subsection{The bounds on the admissible controls}

Definitely, the four bounds on the admissible controls affect the optimal effectiveness of an admissible APT defense strategy. Now, let us examine these influences.

\begin{exm}
Consider a set of problems (16) in which $G$ is the scale-free network generated in Example 1, $T=20$, $\beta = 0.001$, $a_i=0.1$, $0 \leq i \leq N$, and $C_i(0)=0.1, 0 \leq i \leq N$. 
\begin{enumerate}
  \item[(a)] Let $\underline{y} = 0.1$, $\overline{y} =0.7$. Fig. 5(a)-(c) exhibits the influence of $\underline{x}$ and $\overline{x}$ on $OL^*$, $OC^*$, and $OJ^*$, respectively.
  \item[(b)] Let $\underline{x} = 0.1$, $\overline{x} =0.7$.  Fig. 5(d)-(f) displays the influence of $\underline{y}$ and $\overline{y}$ on $OL^*$, $OC^*$, and $OJ^*$, respectively.
\end{enumerate}
\end{exm}

\begin{figure}[H]
	\setlength{\abovecaptionskip}{0.cm}
	\setlength{\belowcaptionskip}{-0.cm}
	\centering
	\includegraphics[width=1\textwidth]{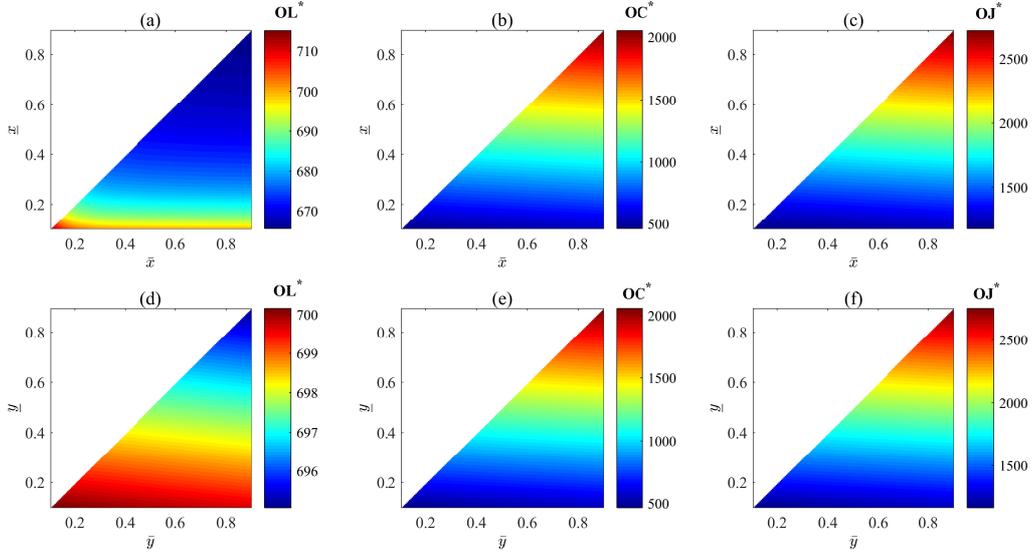}
	\caption{The influence of the four bounds on $OL^*$, $OC^*$, and $OJ^*$ in Example 4.}
\end{figure}

\begin{figure}[H]
	\setlength{\abovecaptionskip}{0.cm}
	\setlength{\belowcaptionskip}{-0.cm}
	\centering
	\includegraphics[width=1\textwidth]{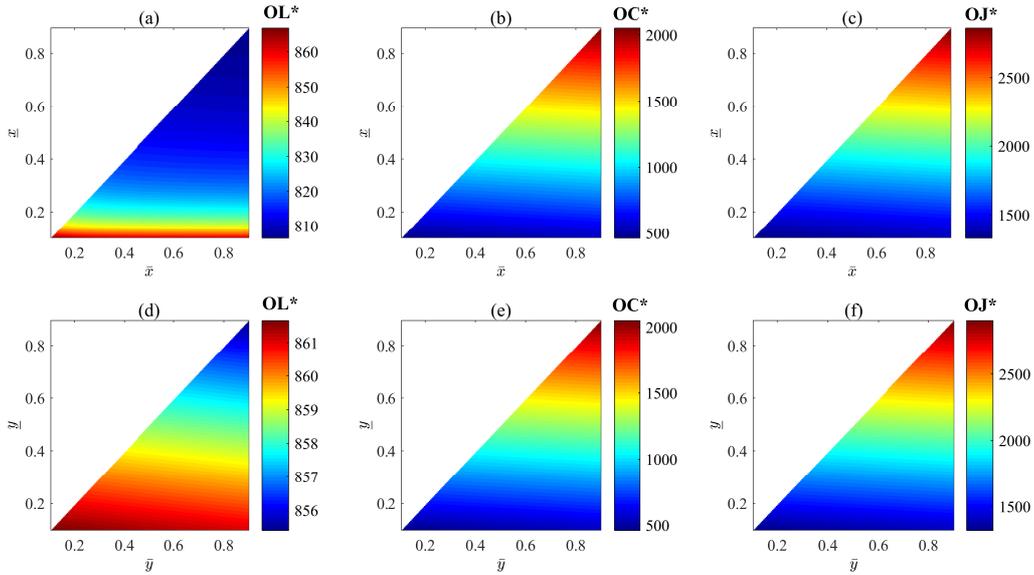}
	\caption{The influence of the four bounds on $OL^*$, $OC^*$, and $OJ^*$ in Example 5.}
\end{figure}

\begin{exm}
Consider a set of problems (16) in which $G$ is the small-world network generated in Example 2, $T=20$, $\beta = 0.001$, $a_i=0.1$, $0 \leq i \leq N$, and $C_i(0)=0.1, 0 \leq i \leq N$. 
\begin{enumerate}
  \item[(a)] Let $\underline{y} = 0.1$, $\overline{y} =0.7$. Fig. 6(a)-(c) exhibits the influence of $\underline{x}$ and $\overline{x}$ on $OL^*$, $OC^*$, and $OJ^*$, respectively.
  \item[(b)] Let $\underline{x} = 0.1$, $\overline{x} =0.7$.  Fig. 6(d)-(f) displays the influence of $\underline{y}$ and $\overline{y}$ on $OL^*$, $OC^*$, and $OJ^*$, respectively.
\end{enumerate}
\end{exm}

\begin{exm}
Consider a set of problem (16) in which $G$ is the realistic network given in Example 3, $T=20$, $\beta = 0.001$, $a_i=0.1$, $0 \leq i \leq N$, and $C_i(0)=0.1, 0 \leq i \leq N$. 
\begin{enumerate}
	\item[(a)] Let $\underline{y} = 0.1$, $\overline{y} =0.7$. Fig. 7(a)-(c) exhibits the influence of $\underline{x}$ and $\overline{x}$ on $OL^*$, $OC^*$, and $OJ^*$, respectively.
	\item[(b)] Let $\underline{x} = 0.1$, $\overline{x} =0.7$.  Fig. 7(d)-(f) displays the influence of $\underline{y}$ and $\overline{y}$ on $OL^*$, $OC^*$, and $OJ^*$, respectively.
\end{enumerate}
\end{exm}

\begin{figure}[H]
	\setlength{\abovecaptionskip}{0.cm}
	\setlength{\belowcaptionskip}{-0.cm}
	\centering
	\includegraphics[width=1\textwidth]{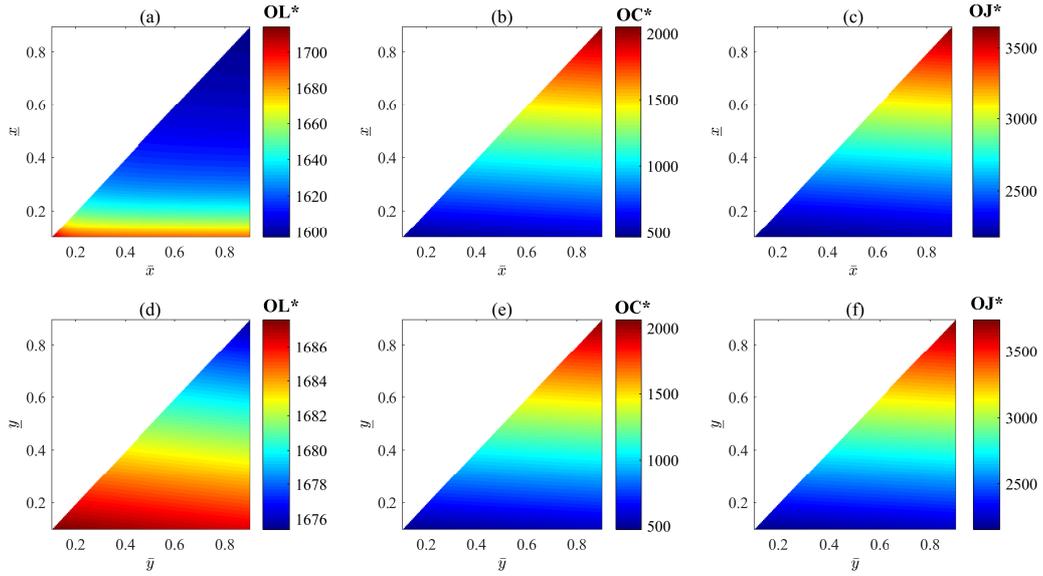}
	\caption{The influence of the four bounds on $OL^*$, $OC^*$, and $OJ^*$ in Example 6.}
\end{figure}

The following conclusions are drawn from the above three examples.

\begin{enumerate}
  
  \item[(a)] With the increase of the two lower bounds, $OL^*$ goes down, but $OC^*$ and $OJ^*$ go up. In practice, the two lower bounds should be chosen carefully so that a balance between the expected loss and the overall cost is achieved.
  
  \item[(b)] The influence of the two upper bounds on $OL^*$, $OC^*$, and $OJ^*$ is almost negligible. 
  
\end{enumerate}

\subsection{The network topology}

Obviously, the topology of the network in an organization affects the optimal effectiveness of an admissible APT defense strategy. Now, let us inspect this influence.

\begin{exm}
Consider a set of problems (16) in which $G \in \{G_i : 1 \leq i \leq 7\}$, where $G_i$ is a scale-free network with $N = 100$ nodes and a power-law exponent of $\gamma_i = 2.7+ 0.1 \times i$, $T = 20$, $\beta = 0.001$, $\underline{x} = \underline{y} = 0.1$, $\overline{x} = \overline{y} = 0.7$, $a_i=0.1$, $0 \leq i \leq N$, and $C_i(0)=0.1, 0 \leq i \leq N$. Figure 8 displays the influence of the power-law exponent on $OL^*$, $OC^*$, and $OJ^*$, respectively. 
\end{exm}

\begin{figure}[H]
	\setlength{\abovecaptionskip}{0.cm}
	\setlength{\belowcaptionskip}{-0.cm}
	\centering
	\includegraphics[width=1\textwidth]{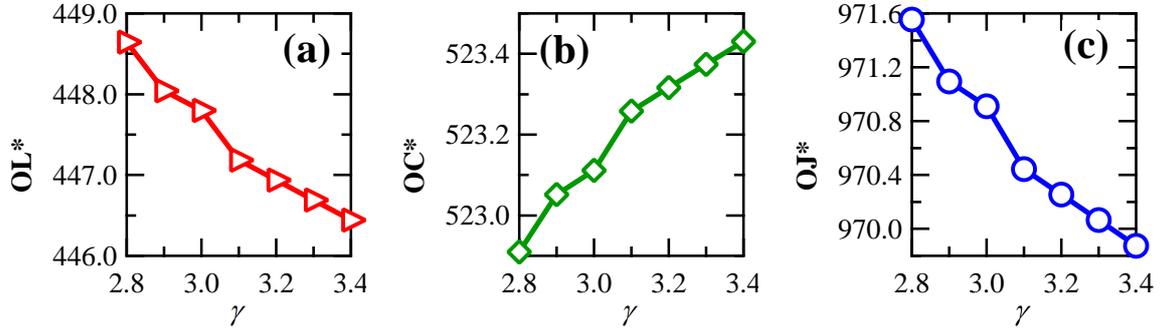}
	\caption{The influence of the power-law exponent of a scale-free network on $OL^*$, $OC^*$, and $OJ^*$ in Example 7.}
\end{figure}

It is seen from this example that, with the increase of the power-law exponent of a scale-free network, $OL^*$ and $OJ^*$ decline, but $OC^*$ inclines. It is well known that the heterogeneity of a scale-free network increases with its power-law exponent. Therefore, a homogeneously mixed access network is better in terms of the optimal defense effectiveness.

\begin{exm}
Consider a set of problems (P) in which $G \in \{G_i : 1 \leq i \leq 5\}$, where $G_i$ is a small-world network with $N = 100$ nodes and an edge-rewiring probability of $p_i = 0.1 \times i$, $T = 20$, $\beta = 0.001$, $\underline{x} = \underline{y} = 0.1$, $\overline{x} = \overline{y} = 0.7$, $a_i=0.1$, $0 \leq i \leq N$, and $C_i(0)=0.1, 0 \leq i \leq N$. Figure 9 exhibits the influence of the edge-rewiring probability on $OL^*$, $OC^*$, and $OJ^*$, respectively. 
\end{exm}

\begin{figure}[H]
	\setlength{\abovecaptionskip}{0.cm}
	\setlength{\belowcaptionskip}{-0.cm}
	\centering
	\includegraphics[width=1\textwidth]{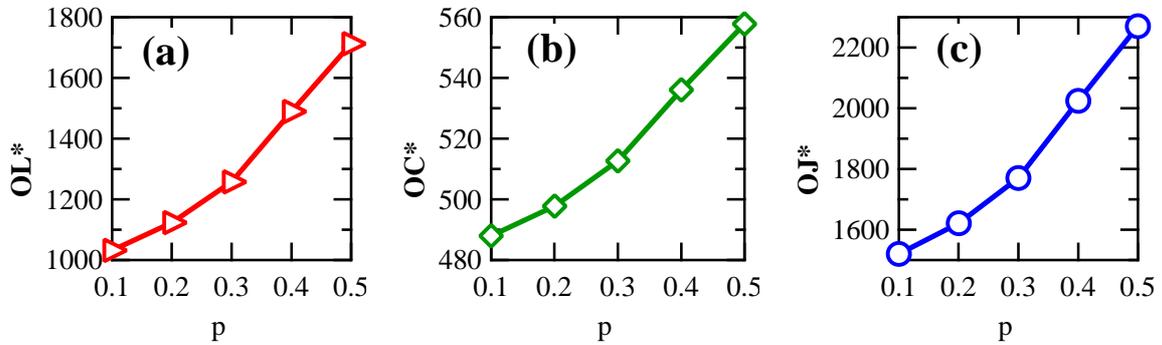}
	\caption{The influence of the edge-rewiring probability of a small-world network on $OL^*$, $OC^*$, and $OJ^*$ in Example 8.}
\end{figure}

It is seen from this example that, with the increase of the randomness of a small-world network, $OL^*$, $OC^*$, and $OJ^*$ rise rapidly. Hence, a randomly connected access network is better from the perspective of the optimal defense effectiveness.

\section{Concluding remarks}

This paper has addressed the APT defense problem, i.e., the problem how to effectively defend against APTs. By introducing an APT attack-defense model and quantifying the effectiveness of an APT defense strategy, we have modeled the APT defense problem as an optimal control problem in which an optimal control represents a most effective APT defense strategy. Through theoretical study, we have presented the optimality system for the optimal control problem. This implies that an optimal control can be derived by solving the optimality system. The influence of some factors on the optimal effectiveness of an APT defense strategy has been examined. 

There are many relevant problems to be resolved. The expected loss and overall cost of an APT defense strategy should be appropriately balanced to adapt to specific application scenarios. In practice, the implementation of a recommended defense strategy needs a great effort; the security level of all the systems in an organization must be labelled accurately \cite{Cole2013}, the defense budget must be made, and the robustness of the defense strategy must be evaluated. As the topology of the access network in an organization may well vary with time, the approach proposed in this work should be adapted to time-varying networks \cite{Schwarzkopf2010, Holme2012, Karsai2014, Valdano2015}. It is of practical importance to deal with the APT defense problem in the game-theoretical framework, where the attacker is strategic \cite{Alpcan2010, Manshaei2013, LiangXN2013, HuPF2015}.

\section*{Acknowledgments}

The authors are grateful to the anonymous reviewers and the editor for their valuable comments and suggestions. This work is supported by National Natural Science Foundation of China (Grant No. 61572006), National Sci-Tech Support Program of China (Grant No. 2015BAF05B03), and Fundamental Research Funds for the Central Universities (Grant No. 106112014CDJZR008823).

%% The Appendices part is started with the command \appendix;
%% appendix sections are then done as normal sections
%% \appendix

%% \section{}
%% \label{}
%\{References}
%%
%% Following citation commands can be used in the body text:
%% Usage of \cite is as follows:
%%   \cite{key}          ==>>  [#]
%%   \cite[chap. 2]{key} ==>>  [#, chap. 2]
%%   \citet{key}         ==>>  Author [#]

%% References with bibTeX database:

%\bibliographystyle{model1-num-names}
\bibliography{<your-bib-database>}

%% Authors are advised to submit their bibtex database files. They are
%% requested to list a bibtex style file in the manuscript if they do
%% not want to use model1-num-names.bst.

%% References without bibTeX database:

\pagebreak

The authors declare that there is no conflict of interest regarding the publication of this paper.

\end{document}